\documentclass[12pt]{article}
\usepackage{a4,amsmath,epsfig,amssymb}
\usepackage[latin1]{inputenc}
\usepackage{graphicx}
\usepackage{caption}
\usepackage{subcaption}
\usepackage{feynmp}
\begin{document}
\bigskip\bigskip

\begin{center}
{\large \bf Searching for New Physics \\
in the Differential Decay Width \\
of a $\Lambda_b$ Semi-leptonic Decay} 
\end{center}
\vspace{8pt}
\begin{center}
\begin{large}

E.~Di Salvo$^{a,b,c,}$\footnote{disalvo@ge.infn.it}
and
Z.~J.~Ajaltouni$^{a,}$\footnote{ziad@clermont.in2p3.fr}
\end{large} 

\bigskip
$^a$ 
Laboratoire de Physique de Clermont - UCA \\
4 Av. Blaise Pascal, TSA60026, F-63178 Aubi\`ere Cedex, France\\

\noindent  
$^b$ 
Dipartimento di Fisica 
 Universit\'a di Genova \\
Via Dodecaneso, 33, 16146 Genova, Italy
 
\noindent  
$^c$
I.N.F.N. - Sez. Genova,\\
Via Dodecaneso, 33, 16146 Genova, Italy \\  

\noindent  

\vskip 1 true cm
\end{center} 
\vspace{1.0cm}

\vspace{6pt}
\begin{center}{\large \bf Abstract}

We propose to investigate the effects of new physics in the semi-leptonic sequential decay $\Lambda_b \to \Lambda_c (\to \Lambda \pi) \tau{\bar \nu}_{\tau}$. Firstly, we write the general, model independent, non-covariant expression of the differential decay width of the process.
Then, we calculate that observable according to three different types of new physics interactions, which might explain the tension of data with the standard model predictions. We find that some coefficients of the differential decay width are sensitive to the kind of interaction that is assumed. The measurements that we suggest seem to be feasible.

\end{center}

\vspace{10pt}

\centerline{PACS numbers: 13.30.Ce, 12.15.-y, 12.60.-i}

\vspace{10pt}

\centerline{Keywords: New Physics, Semi-leptonic decays}

\newpage

\section{Introduction}
Hints to physics beyond the standard model (SM) are given by various data about the semi-leptonic decays $B \to D^{(*)} \tau \bar{\nu}_{\tau}$, which have exhibited strong tensions with the SM predictions\cite{bb}. 
Also some data on the $B\to K^* \ell^+\ell^-$ decays show discrepancies with the SM\cite{bfn}, but recent LHCb results\cite{lh1,lh2} agree with muon-electron universality. As regards the basic process
\begin{equation}
b\to c \tau \bar{\nu}_{\tau}, \label{ccu}
\end{equation}
further tensions with the SM are found in decays of the type $B_c\to J/\psi (\eta_c) \tau \bar{\nu}_{\tau}$\cite{alb,dub,iss}. Incidentally, the data on $B_c\to J/\psi (\eta_c) \pi$ decay are in agreement with the SM\cite{iss}, whence one could infer that, unlike leptons, hadrons are not sizably affected by new physics (NP).
 
All that leads to focusing on decays which involve the charged current (\ref{ccu}): more and definitive confirmations of NP are searched for in semi-leptonic decays of heavy baryons, like $\Omega_b$, $\Sigma_b$ and $\Lambda_b$ [8-18]. In particular, a test of lepton flavor universality has been performed by LHCb\cite{cern} on the $\Lambda_b$ decay to $\Lambda_c \ell \bar{\nu}_{\ell}$.

Moreover, recently, for those decays where spinning particles are involved, measurements concerning asymmetries, polarizations and differential decay widths have been proposed[1,5,8,9,20-31] and realized\cite{Be1,Be2}.
 
The aim of the present letter is to study the differential decay width of the baryonic cascade decay 
\begin{equation}
\Lambda_b \to \Lambda_c \tau^- \bar{\nu}_{\tau}\to (\Lambda \pi) \tau^- \bar{\nu}_{\tau}.  \label{slh} 
\end{equation}

In particular, we use the non-covariant formalism for parametrizing the
differential decay width, then we adopt a simple model of form factor for 
calculating the coefficients of this distribution, according to three different scenarios of NP:

- a pure left-handed, $L$ = $V-A$ interaction;

- a right-handed, $R$ = $V+A$ interaction;

- charged-Higgs exchange, $H$ = $S-P$.

The sequential decays of spinning particles, like (\ref{slh})\cite{fs}, allow to determine, in principle, the longitudinal polarization of the intermediate baryon, say $\Lambda_c$, and the interference between different hadron helicity amplitudes, which is problematic in the case of $B \to D^{(*)} \tau \bar{\nu}_{\tau}$ decays. 
As we shall see, such observables may help discriminating among different interactions. Indeed, although the three scenarios above do not cover all possible kinds of NP which may explain the tensions of data with the SM[16,21,23,34-37], they help illustrating the sensitivity of the distribution to the different interactions. 
 
In this letter, firstly, we parametrize the differential decay width of the process (\ref{slh}); then, we calculate the coefficients of the parametrization according to the above mentioned interactions.

\section{Differential Decay Width}

We consider the (normalized) differential decay width of the decay (\ref{slh}), with the $\tau$ lepton unobserved. It is defined as

\begin{equation}
I(\Omega_{\Lambda_c},\Omega_{\Lambda};Q) = \frac{1}{\Gamma}\frac{d\Gamma}{d\Phi}, \ ~~~~ \ d\Phi=\frac{1}{2^4(2\pi)^7}\frac{p p_l p_{\Lambda}}{m_{\Lambda_b}m_{\Lambda_c}} dQ d\Omega_{\Lambda_c} 
d\Omega_{\Lambda}. \label{mf}
\end{equation}

Here ($p$, $\Omega_{\Lambda_c}$) and ($p_{\Lambda}$, $\Omega_{\Lambda}$) denote the momenta and directions, respectively, of the  $\Lambda_c$ in the $\Lambda_b$ rest frame (RF) and of the ${\Lambda}$ in the $\Lambda_c$ RF; 
$\Omega_{\Lambda_c} \equiv (\theta_{\Lambda_c},\phi_{\Lambda_c})$
and $\theta_{\Lambda_c}$ and $\phi_{\Lambda_c}$ are, respectively, the polar and azimuthal angle of the momentum of $\Lambda$; an analogous notation holds for $\Omega_{\Lambda}$.

The general structure of the Dirac operators entails that the $\tau$-${\bar \nu}_{\tau}$ system carries spin 0 or 1: we denote by ${\cal B}^*$ the intermediate virtual boson ($W^*$, $H^*$, {\it etc}.). 

We choose, for the decay $\Lambda_b \to \Lambda_c {\cal B}^*$, the canonical quantization axis along the polarization vector $\vec{P}_{\Lambda_b}$ of the parent resonance. Moreover, we integrate the differential decay width over the effective mass of the $\tau$-${\bar \nu}_{\tau}$ system, denoted by $Q$. Therefore, according to the Jacob-Wick formalism\cite{jw}, we get[39-41]
\begin{eqnarray}
I(\Omega_{\Lambda_c},\Omega_{\Lambda}) &=& \frac{1}{16\pi^2}
\{1+ \beta_L^{\Lambda} \beta_L^{\Lambda_c} \cos\theta_{\Lambda} + P^{\Lambda_b} [\cos\theta_{\Lambda_c}(\beta_L^{\Lambda_c}+ \beta_L^\Lambda \cos\theta_{\Lambda})+
\nonumber
\\
&+& 4 \beta_L^\Lambda \sin\theta_{\Lambda_c}\sin\theta_{\Lambda} (\beta_T^{\Lambda_c} \cos\phi_{\Lambda} + \beta_N^{\Lambda_c}\sin\phi_{\Lambda})]\}. \label{ddw}
\end{eqnarray}
Here $P^{\Lambda_b}$ = $|\vec{P}_{\Lambda_b}|$,
\begin{eqnarray}
\beta_L^\Lambda &=& |b_+|^2-|b_-|^2, \ ~~~~ \ ~~~~ \ ~~~~ \ ~~~ \ ~~~  ~~~ \
\\
\beta_L^{\Lambda_c}  &=& {\cal N}^{-1}\int_{m_l}^{M_-}  dQ p p_l 
(\Delta K_S+\Delta K_L+\Delta K_U),
\\
\beta_{T,N}^{\Lambda_c} &=&  {\cal N}^{-1}\int_{m_l}^{M_-}  dQ p p_l \alpha_{T,N}^{\Lambda_c},
\\
{\cal N} &=& \int_{m_l}^{M_-} dQ p p_l (K_S+K_L+K_U), ~~~ \ ~~~  ~~~ \
\label{nrm}
\end{eqnarray}
$M_-$ = $m_{\Lambda_b}$ - $m_{\Lambda_c}$ and $m_l$ is the mass of the $\tau$-lepton. Moreover,
\begin{eqnarray}
K_S(\Delta K_S) &=& \sum_{r,r'}(H^{0rr'}_{0++}\pm H^{0rr'}_{0--}) L^{0rr'}_0, \ ~~~~ \ \ ~~~~ \ \ ~~~~ \ \ ~~~~ \  
\\ 
K_L(\Delta K_L) &=& \sum_{r,r'}(H^{1rr'}_{0++}\pm H^{1rr'}_{0--})(L^{1rr'}_0+L^{1rr'}_{-1}), \ ~~~~ \ \ ~~~~ \ \ ~~~~ \ \ ~~~~ \ 
\\ 
K_U(\Delta K_U) &=& \sum_{r,r'}(H^{1rr'}_{1++}\pm H^{1rr'}_{-1--})(L^{1rr'}_0+L^{1rr'}_{-1}), \ ~~~~ \ \ ~~~~ \ \ ~~~~ \ \ ~~~~ \ 
\\
\alpha_{T,N}^{\Lambda_c}  &=& 2\Re(-2\Im)\sum_{r,r'}[H^{0rr'}_{0+-} L^{0rr'}_0 + H^{1rr'}_{0+-}(L^{1rr'}_0 + L^{1rr'}_{-1})],
\\
H^{Jrr'}_{\mu\lambda\lambda'} &=&  a^{Jr}_{\mu \lambda} a^{Jr'*}_{\mu\lambda'} \ ~~~~ \
 L^{Jrr'}_{\eta} = \frac{1}{2J+1} c^{Jr}_{\lambda_l+} c^{Jr'*}_{\lambda_l+}, \ ~~~~ \ \eta=\lambda_l-1/2.
\end{eqnarray}
$a^{Jr}_{\mu\lambda}$, $c^{Jr}_{\lambda_l+}$ and $b_{\beta}$ are respectively the helicity amplitudes of the decays $\Lambda_b \to \Lambda_c {\cal B}^*$, ${\cal B}^*\to \tau^- \bar{\nu}_{\tau}$ and $\Lambda_c \to \Lambda \pi$; $\mu$, $\lambda$ and $\lambda_l$ denote, respectively, the helicities of ${\cal B}^*$, $\Lambda_c$ and $\tau$, while the index $r$ refers to the type of interaction considered ($L$, $R$, $H$). Last, we normalize the coefficients $b_{\beta}$ so that
\begin{equation}
|b_+|^2+|b_-|^2 = 1. 
 \label{ncst}
\end{equation}
Note that the amplitudes $a^{Jr}_{\mu\lambda}$ and $c^{Jr}_{\lambda_l+}$ depend on $Q$. 

We conclude this section with a remark. The phase of the product 
$a^{Jr}_{\mu \lambda} a^{Jr'*}_{\mu\lambda'}$ is $T$-odd, as shown in ref. 42. Then also the imaginary part of $H^{Jrr'}_{\mu\lambda\lambda'}$ is $T$-odd. But this feature is shared also by $\beta_N^{\Lambda_c}$, because $L^{Jrr'}_{\eta}$ is real, as we shall see in the next section.
Therefore, since, in the decay $\Lambda_b \to \Lambda_c \tau^- \bar{\nu}_{\tau}$, the final-state interactions are negligibly small, a non-zero value of $\beta_N^{\Lambda_c}$ would imply time reversal violation. In this connection, for the decay $\Lambda_b \to \Lambda_c \mu^- \bar{\nu}_{\mu}$, a preliminary determination of this observable was made a few years ago\cite{ko,ak}.

\section{NP Interactions}   

The decay amplitude for the process considered reads as
\begin{equation}
A = G'(T_{SM}+ z T_{NP}), ~~~ z = x e^{i\psi}, ~~~
G' = V_{bc} \frac{G}{\sqrt{2}}. \label{mampl}
\end{equation}
Here $G$ is the Fermi constant, $V_{bc}$ the Cabibbo-Kobayashi-Maskawa matrix element for the quark transition $b \to c$ and $T_{SM}$ and $z T_{NP}$ take account of, respectively, the SM and NP interaction, with $x$ real and positive. In particular,
\begin{equation}
T_{SM} = h^\alpha_L g_{\alpha\beta} l^{L\beta}, \label{smal}
\end{equation}
where $h^\alpha_L$ and $l_\alpha^L$ describe respectively the hadronic and leptonic vertex, {\it i. e.}, 
\begin{equation}
h^\alpha_L = {\bar u}_{\Lambda_c} (\Gamma^{\alpha}-\Gamma^{\alpha}_5) u_{\Lambda_b} ~~~ \mathrm{and} ~~~ l_\alpha^L = {\bar u}_l \gamma_{\alpha} v_{\bar{\nu}}. \label{SM}
\end{equation}
Here
\begin{equation}
\Gamma^{\alpha} = {\cal F}_1\gamma^{\alpha}+{\cal F}_2i\sigma^{\alpha\beta}q_{\beta} +{\cal F}_3q^{\alpha},
\ ~~~~ \
\Gamma^{\alpha}_5= [{\cal G}_1\gamma^{\alpha}+{\cal G}_2 i\sigma^{\alpha\beta}q_{\beta} +{\cal G}_3 q^{\alpha}]\gamma_5 \label{pscl}
\end{equation}
and the form factors ${\cal F}_i$ and ${\cal G}_i$ are functions of $Q^2$.

As regards NP, 

- the $L$-interaction amounts to a re-scaling of the SM amplitude, which reads as
\begin{equation}
T^L_{NP} = T_{SM} 
\end{equation}
and the possible complex phase induced by $z$ in Eq. (\ref{mampl}) is ruled out, as unphysical;

- the $R$-interaction is obtained from the SM expression by changing the sign of $\Gamma^{\alpha}_5$ in the first Eq. (\ref{SM}):
\begin{equation}
T^R_{NP} = {\bar u}_{\Lambda_c} (\Gamma^{\alpha} + \Gamma^{\alpha}_5) u_{\Lambda_b} l_\alpha^L; 
\end{equation}

- last, the $H$-interaction reads as 

\begin{equation}
T^H_{NP} = {\bar u}_{\Lambda_c} ({\cal F}_0 - \gamma_5 {\cal G}_0) u_{\Lambda_b}{\bar u}_l v_{\bar{\nu}}, \label{dda}
\end{equation}
where ${\cal F}_0$ and ${\cal G}_0$ are the scalar form factors. These can be related to the previous form factors thanks to the equations of motion\cite{dsfa}: 
\begin{equation}
{\cal F}_0 = {\cal F}_1\rho_- + {\cal F}_3\frac{Q^2}{m_b-m_c}, ~~~~
{\cal G}_0 = {\cal G}_1\rho_+ - {\cal G}_3\frac{Q^2}{m_b+m_c};
\end{equation}
Here 
\begin{equation}
\rho_{\pm} = \frac{m_{\Lambda_b} \pm m_{\Lambda_c}}{m_b \pm m_c} 
\end{equation}
and $m_b$ and $m_c$ are the masses of the $b$- and $c$-quark respectively, $m_b$ = 4.18 $GeV$ and $m_c$ = 1.28 $GeV$.

The expressions above are considerably simplified, since we adopt the Isgur-Wise (IW\cite{iw}) form factor for heavy-to-heavy  transitions\cite{rhk,blr}, {\it i. e.},
\begin{equation}
{\cal F}_2 = {\cal F}_3 = {\cal G}_2 = {\cal G}_3 = 0, ~~~~
{\cal F}_1 = {\cal G}_1 = \zeta. \label{iw0}
\end{equation}

The helicity amplitudes for vector interactions are deduced in refs. 11, 14 and 48, by introducing four mutually orthogonal unit four-vectors in the Minkowski space-time. The completeness relation for such four-vectors is inserted into Eq. (\ref{smal}), allowing to separate the leptonic amplitude from the hadronic one and to extract the helicity amplitudes. For the $L$-interaction, one has
\begin{eqnarray}
a^{0L}_{0\lambda} &=& \zeta \frac{F}{Q} [p_0+\chi p -2\lambda(p+\chi p_0)], ~~~~  a^{1L}_{0\lambda} = \frac{F}{Q} [p+\chi p_0 -2\lambda(p_0+\chi p)],
\\
a^{1L}_{+1+} &=& \sqrt{2}\zeta F(1-\chi), ~~~~ \ ~~~~ \ ~~~~ \ ~~~~
a^{1L}_{-1-} =  -\sqrt{2}\zeta F(1+\chi) 
\end{eqnarray}
and
\begin{equation}
c^{0L}_{++} = -c^{1L}_{++} = F_l(1-\chi_l), ~~~~  
c^{1L}_{-+} = \sqrt{2} F_l(1+\chi_l). \label{hhc}
\end{equation}
Here 
\begin{eqnarray}
F &=& \sqrt{2m_{\Lambda_b}(E_{\Lambda_c}+m_{\Lambda_c})}, ~~~ \chi = \frac{p}{E_{\Lambda_c}+m_{\Lambda_c}}, ~~~ E_{\Lambda_c} = \sqrt{p^2+m_{\Lambda_c}^2},
\\
F_l &=& \sqrt{p_l(E_l+m_l)}, ~~~ ~~~ \chi_l = \frac{p_l}{E_l +m_l}, ~~~
~~~ E_l = \sqrt{p_l^2+m_l^2}.
\end{eqnarray}

As far as the $R$-amplitudes are concerned, we have 
\begin{equation}
a^{1R}_{\mu \lambda} = a^{1L}_{-\mu -\lambda},
\end{equation}
while the leptonic amplitudes are still given by Eqs. (\ref{hhc}).

Last, as regards the $H$-interaction, it is straightforward to deduce
\begin{equation}
a^{0H}_{0\lambda} = \zeta F (\rho_- + 2\lambda \chi\rho_+), ~~~~
c^{0H}_{++} = F_l(1+\chi_l). \label{Hha}
\end{equation}

\section{Numerical Results - Discussion}

Now we calculate, according to each interaction, the observables $\beta_L^{\Lambda_c}$, $\beta_T^{\Lambda_c}$ and $\beta_N^{\Lambda_c}$, which appear in Eq. (\ref{ddw}). To this end, we consider two different parametrizations for the Isgur-Wise form factor, {\it i. e.},
\begin{eqnarray}
\zeta_1 &=& 1-\rho^2[\omega-1]+1/2\sigma^2[\omega-1]^2,
\\
\zeta_2 &=& [2/(\omega+1)]^{\alpha}, ~~~ \alpha=3.5+1.2/\omega\cite{rhk},
\end{eqnarray} 
with
\begin{equation}
\omega =
\frac{m_{\Lambda_b}^2+m_{\Lambda_c}^2-Q^2}{2m_{\Lambda_b}m_{\Lambda_c}}.
\end{equation}
As regards $\zeta_1$, we adopt two different sets of values of the parameters:

a) $\rho^2=1.47$, ~~~ $\sigma^2 = 1.90$\cite{klw},

b) $\rho^2=2.04\pm0.08$, ~~~ $\sigma^2 = 3.16\pm0.38$\cite{blr};

Moreover, in order to fix the relative strength $x$ and the phase $\psi$ of the NP interaction, we use the bounds imposed by the data on the $B$ semi-leptonic decays, according to the last $HFLAV$ analysis\cite{hflav}. In particular, we infer from such an analysis the value of a parameter $\Delta$, which is defined as
\begin{equation}
1+\Delta =  \frac{\Gamma}{\Gamma_{SM}} = 
\frac{{\cal R}(\Lambda_c)}{{\cal R}_{SM}(\Lambda_c)}, ~~~~
{\cal R}(\Lambda_c)= \frac{\Gamma}{\Gamma_{\mu}}. 
\end{equation}
Here $\Gamma$ is the partial width of the decay $\Lambda_b \to \Lambda_c \tau^- \bar{\nu}_{\tau}$ and $\Gamma_{\mu}$ the one for $\Lambda_b \to \Lambda_c \mu^- \bar{\nu}_{\mu}$, the subscript $SM$ denoting the SM prediction of the observable. 

In this connection, we observe that ${\cal R}_{SM}$ can be calculated according to our model, {i. e.},
\begin{equation}
{\cal R}_{SM} = \frac{\cal{N}}{\cal{N}_{\mu}},
\end{equation}
where ${\cal N}_{\mu}$ is obtained from Eq. (\ref{nrm}) by substituting $m_l$ $\to$ $m_{\mu}$. The result is 
\begin{equation}
{\cal R}_{SM}(\Lambda_c) = 0.330\pm0.018,
\end{equation}
which is consistent with the value obtained by Bernlochner {\it et al.}\cite{bnl} $0.324\pm0.004$. The experimental value found by the LHCb collaboration\cite{cern}, 
\begin{equation}
{\cal R}(\Lambda_c) = 0.242\pm 0.076,
\end{equation}
yields 
\begin{equation}
\Delta = -0.201\pm0.251. \label{lmb}
\end{equation}
The $HFLAV$ analysis of $B$ semi-leptonic data provides a more precise evaluation of the parameter, especially about the decays $ B \to D^* \ell \bar{\nu}_{\ell}$; indeed,
\begin{equation}
{\cal R}(D^*) = 0.295 \pm 0.011 \pm 0.008, ~~~~ {\cal R}_{SM}(D^*) = 0.258\pm 0.005
\end{equation}
and therefore
\begin{equation}
\Delta = 0.143^{+0.078}_{-0.075}. \label{err}
\end{equation}
We use this value - which is compatible with (\ref{lmb}) to within $\sim 1.1\sigma$ - for determining the coupling, $z$ = $x$ $e^{i\psi}$, taking into account, for each kind of interaction, the bounds established by the analyses on the $B$ decays[52-56].

- The $L$-interaction sets no limits on $\psi$, then, choosing it to be zero, we get 
\begin{equation}
x_L = \sqrt{1+\Delta}-1 = 0.069\pm0.036.
\end{equation}

- The $R$-interaction demands, according to the previous analyses[53-56], $\psi$ = $\pi/2$, whence
\begin{equation}
x_R = \sqrt{\Delta} = 0.378^{+0.092}_{-0.117}.
\end{equation}

- As regards the $H$-interaction, $x$ and $\psi$ are restricted to the intersection of two circular crowns in the complex plane of the coupling. Indeed, a circumference is characterized, with respect to these variables,  by the equation
\begin{equation}
r_0 x^2 + r_1 x \cos\psi = \Delta. \label{xpsi}
\end{equation}
Here, according to our analysis, $r_0$ = 0.784 and $r_1$ = 0.712, while the $B$-data entail\cite{ivn2} $r_0$ = 0.635 and $r_1$ = 0.965, together with the inequality $x \cos\psi$ $\leq$ -0.76; the uncertainties on $\Delta$ and on the form factors determine the radii of the circular crowns. Such bounds imply $\psi$ $\simeq$ $2.274$, while $x$ is the corresponding positive root of Eqs. (\ref{xpsi}), with $r_0$ and $r_1$ fixed according to our calculations: it results to be $x = 0.872^{+0.109}_{-0.105}$.

The results of our analysis are summarized in Table \ref{tab:one}. 
All of our predictions are affected by a systematic error, due to the uncertainty on the form factor. Some of them are affected also by a statistical error, which, in such cases, is reported firstly on the table. 
 
Eq. (\ref{ddw}) implies that the effective observables, which can be inferred from data of the differential decay width, consist of the products 
\begin{equation}
{\cal P}_L = \beta_L^{\Lambda} \beta_L^{\Lambda_c} \ ~~~~ \ \mathrm{and} \ ~~~~ \  {\cal P}_{T(N)} = \beta_L^{\Lambda}\beta_{T(N)}^{\Lambda_c} P^{\Lambda_b}. 
\end{equation}
Then, since $\beta_L^\Lambda$ = $-0.84\pm 0.09$\cite{pdg}, the observable 
${\cal P}_L$ allows to distinguish between the $L$-interaction and the other two. On the contrary, ${\cal P}_T$ and ${\cal P}_N$ - the latter of which is quite sensitive to the type of NP interaction - depend crucially on $P^{\Lambda_b}$, which is more sizable in reactions of the type $e^+ - e^-$ $\to$ $Z^*$ $\to$ $b\bar{b}$[58-60] than in $p-p$ collisions\cite{cms}.

At this point, it is worth spending a few words on the experimental feasibility of the measurement that we propose. In fact, detecting the decay (\ref{slh}) involves at least two neutrinos, which makes the event reconstruction quite difficult. However, what we demand for our proposal, is not the momentum and energy of each $\tau$, since we have integrated over $Q$ and over the solid angle of the lepton. In particular, at the LHC  energy, the two neutrinos emitted by the $\Lambda_b$ are almost collinear, which makes the missing energy-momentum reconstruction  a realistic task.

To conclude, our study indicates that it is possible - through a realistically feasible measurement - to discriminate among various NP candidate interactions. Moreover, the $R$- and $H$-interaction, where a non-trivial phase is involved,  would imply time reversal violation, according to the remark at the end of Sect. 2. Last, since the Dirac matrices impose restrictions on the angular momentum of the $\tau$-${\bar \nu}_{\tau}$ system, an angular analysis of the $\tau$ distribution, if realizable, could allow to detect experimentally possible spurious events.

\vskip 0.25cm

\centerline{\bf Acknowledgments}
The authors are deeply indebted to their friend S. Monteil for stimulating discussions. One of them (EDS) is also grateful to his friend F. Fontanelli for useful remarks.

\begin{table*}
\begin{center}
\caption{Parameters and predictions of observables according to the different interactions.}
\begin{tabular}{|c|c|c|c}
\hline\hline
$~~~~~~~~~~$ & $\beta_L^{\Lambda_c}$ &$\beta_T^{\Lambda_c}$& $\beta_N^{\Lambda_c}$ \\
\hline\hline
$~~~~L~~~~$ & $-0.643\pm0.003$ & $-0.100\pm0.001$       & 0.0       
\\
$~~~~R~~~~$ &$-0.482^{+0.074}_{-0.072}\pm0.002$ & $-0.100\pm0.001 $      & $-0.092^{+0.019}_{-0.018}\pm0.000$  
\\
$~~~~H~~~~$ & $-0.416^{+0.057}_{-0.052}\pm0.003$ & $-0.136^{+0.024}_{-0.032}\pm0.001$ & $-0.217^{+0.017}_{-0.019}\pm0.001$  \\
\end{tabular}
\label{tab:one}       
\end{center}
\end{table*}

\end{document}